\documentstyle[12pt,aaspp4,psfig,side]{article}   
\tighten
\newbox\grsign \setbox\grsign=\hbox{$>$} 
\newdimen\grdimen \grdimen=\ht\grsign
\newbox\laxbox \newbox\gaxbox
\setbox\gaxbox=\hbox{\raise.5ex\hbox{$>$}\llap
     {\lower.5ex\hbox{$\sim$}}}\ht1=\grdimen\dp1=0pt
\setbox\laxbox=\hbox{\raise.5ex\hbox{$<$}\llap
     {\lower.5ex\hbox{$\sim$}}}\ht2=\grdimen\dp2=0pt
\def\gax{\mathrel{\copy\gaxbox}}

\def\g123{GRB~990123}

\begin{document}

\title{Spectral Energy Distributions and Light Curves of
GRB~990123 and its Afterglow}
\author{T.J. Galama$^{1}$,
M.S. Briggs$^{2}$,
R.A.M.J. Wijers$^{3}$,
P.M. Vreeswijk$^{1}$, 
E. Rol$^{1}$,
D. Band$^{4}$,
J. van Paradijs$^{1,2}$,
C. Kouveliotou$^{5,6}$,
R.D. Preece$^{2}$,
M. Bremer$^{7}$,
I.A.  Smith$^{8}$,
R.P.J. Tilanus$^{9}$,
A.G. de Bruyn$^{10,11}$,
R.G. Strom$^{1,10}$,
G. Pooley$^{12}$,
A.J. Castro-Tirado$^{13,14}$,
N. Tanvir$^{15,16}$,
C. Robinson$^{17}$,
K. Hurley$^{18}$,
J. Heise$^{19}$,
J. Telting$^{20}$,
R.G.M. Rutten$^{20}$,
C. Packham$^{20}$,
R. Swaters$^{11}$,
J.K. Davies$^{9}$,
A. Fassia$^{21}$,
S.F. Green$^{22}$,
M.J. Foster$^{22}$,
R. Sagar$^{23}$,
A.K. Pandey$^{23}$,
Nilakshi$^{23}$,
R.K.S. Yadav$^{23}$,
E.O. Ofek$^{24}$,
E. Leibowitz$^{24}$,
P. Ibbetson$^{24}$,
J. Rhoads$^{25}$,
E. Falco$^{26}$,
C. Petry$^{27}$,
C. Impey$^{27}$,
T.R. Geballe$^{28}$,
D. Bhattacharya$^{29}$
}

1. {Astronomical Institute `Anton Pannekoek', University
of Amsterdam, \& Center for High Energy Astrophysics, Kruislaan 403,
1098 SJ Amsterdam, The Netherlands} \\
2. {Physics
Department, University of Alabama in Huntsville, Huntsville AL 35899,
USA} \\
3. {Department of Physics and Astronomy, SUNY Stony
Brook, NY 11794-3800, USA} \\
4. {CASS, University of
California in San Diego, La Jolla, CA 92093, USA}\\
5. {Universities Space Research Association}\\
6. {NASA/MSFC, Code ES-84, Huntsville AL 35812, USA}\\
7. {Institut de Radio Astronomie
Millim\'{e}trique, 300 rue de la Piscine, F--38406 Saint-Martin
d'H\`{e}res, France} \\
8. {Department of Space Physics and Astronomy, Rice
University, MS-108, 6100 South Main, Houston, TX 77005-1892 USA}\\
9. {Joint Astronomy Centre, 660 North A'ohoku Place,
University Park, Hilo, HI 96720, USA}\\
10. {NFRA, Postbus
2, 7990 AA Dwingeloo, The Netherlands} \\
11. {Kapteyn
Astronomical Institute, Postbus 800, 9700 AV, Groningen, The
Netherlands} \\
12. {Mullard Radio Astronomy Observatory,
Cavendish Laboratory, University of Cambridge, Madingley Road
Cambridge CB3 0HE, UK} \\
13. {Laboratorio de Astrof{\'\i}sica
Espacial y F{\'\i}sica Fundamental (LAEFF-INTA), P.O. Box 50727,
E-28080 Madrid, Spain} \\
14. {Instituto de Astrof{\'\i}sica
de Andaluc{\'\i}a (IAA-CSIC), P.O. Box 03004, E-18080 Granada, Spain}\\
15. {Institute of Astronomy, Madingley Road, Cambridge
CB3 0HA, UK} \\
16. {Department of Physical Sciences, University of
Hertfordshire, College Lane, Hatfield, Herts AL10 9AB, UK}\\
17. {National Science Foundation}\\
18. {University of California at Berkeley,
Space Sciences Laboratory, Berkeley, CA, USA 94720-7450}\\
19. {Space Research Organisation Netherlands (SRON),
Sorbonnelaan 2, 3584 CA Utrecht, The Netherlands }\\
20. {Isaac Newton Group, Apartado de Correos, 321, 38780
Santa Cruz de La Palma, Islas Canarias, Spain}\\
21. {Astrophysics Group, Blackett Laboratory, Imperial
College, Prince Consort Road, London SW7 2BZ}\\
22. {Unit for Space Sciences and Astrophysics, School of
Physical 
Sciences, Physics Laboratory, University of Kent at Canterbury,
Canterbury, 
Kent, CT2 7NR, UK} \\
23. {U.P. State Observatory, Nainital}\\
24. {Wise Observatory, Tel Aviv University, Ramat Aviv,
Tel Aviv 699 78, Israel}\\
25. {Kitt Peak National
Observatory, 950 N. Cherry Avenue, PO.O. Box 26732, Tucson, AZ 85726 USA}\\
26. {Harvard-Smithsonian Center for
Astrophysics, Cambridge, MA 02138, USA}\\
27. {Steward Observatory, University of Arizona, Tucson
AZ 85719, USA}\\
28. {Gemini Observatory, 670 N. A'ohoku Place,
University Park, Hilo, HI 96720 USA}\\
29. {Raman Research Institute, Bangalore 560 080, India}

\newpage
{\it This manuscript has been accepted for publication in
Nature. The article is under embargo until publication. We do
place restrictions on any dissemination in the popular media.  You are
free to refer to this paper in your own publications.  For further
enquiries, please contact Titus Galama (titus@astro.uva.nl) or Michael
Briggs (briggs@gibson.msfc.nasa.gov). }

{\bf Gamma-ray bursts (GRBs) are thought to result from the
interaction of an extremely relativistic outflow interacting with a
small amount of material surrounding the site of the explosion.
Multi-wavelength observations covering the $\gamma$-ray to radio
wavebands allow investigations of this `fireball' model. On 23 January
1999 optical emission was detected while the $\gamma$-ray burst was
still underway. Here we report the results of $\gamma$-ray,
optical/infra-red, sub-mm, mm and radio observations of this burst and
its afterglow, which indicate that the prompt and afterglow emissions
from GRB\,990123 are associated with three distinct regions in the
fireball. The afterglow one day after the burst has a much lower peak
frequency than those of previous bursts; this explains the short-lived
nature of the radio emission, which is not expected to reappear. We
suggest that such differences reflect variations in the magnetic-field
strengths in the afterglow emitting regions.}

\section{INTRODUCTION}

The current `industry standard' model for $\gamma$-ray bursts and
their afterglows is the fireball-plus-blastwave model (see Ref. 1 for
a review). It invokes the release of a large amount of energy, of
order M$_\odot c^2$, into a volume less than 1 light-msec across. This
leads to a `fireball' that expands ultra-relativistically (Lorentz
factor $\Gamma \gax 300$); when it runs into the surrounding medium a
`forward shock' ploughs into the medium and heats it, and a `reverse
shock' does the same to the ejecta.  The $\gamma$-ray burst itself is
thought to owe its multi-peaked light curve to a series of `internal
shocks' that develop in the relativistic ejecta before they collide
with the ambient medium.  The heated gas is presumed to form
relativistic electrons and magnetic fields with energy densities near
equipartition, i.e., similar to that of the gas, and then emit
synchrotron radiation$^{2}$.  As the forward shock is weighed down by
increasing amounts of swept-up material it becomes less relativistic,
and produces a slowly fading `afterglow' of X rays, then
UV/optical/IR, and then mm and radio radiation.

M\'esz\'aros \& Rees$^{3}$ and Sari \& Piran$^{4}$ pointed out that
when the external shock first forms, optical emission of $\sim$ 9-14th
magnitude from a reverse shock will be visible during or soon after
the high-energy burst produced by the internal shocks. Other models
have been made which lead to an optical brightness of $\sim$ 18th
magnitude during the $\gamma$-ray burst$^{5}$.  Multi-wavelength
observations during and immediately after the burst are necessary to
detect the emissions from the different regions postulated by these
models and transitions between them.  

The X-ray emission observed with
BeppoSAX during the tails of $\gamma$-ray bursts is consistent with
extrapolations backward in time of the X-ray afterglow detected many
hours later$^{6-10}$, suggesting that the prompt X-ray emission merges
smoothly into the afterglow.  On the other hand, HEAO-1 observed
serendipitously a prompt X-ray signal in GRB\,780506 which disappeared
and then reappeared after a few minutes, suggesting a gap between the
prompt emission and the afterglow$^{11}$.

The ROTSE detection$^{12}$ of prompt optical emission during and
immediately after \g123\ shows that optical observations can address
the question of which physical region is radiating when. The first
ROTSE exposure occurred during the main peak of the burst; weaker
$\gamma$-ray emission was also present during the second and
third ROTSE exposures, but was no longer detected in subsequent
exposures.

We here present an extensive set of multi-wavelength observations for
GRB~990123, which cover the $\gamma$-ray to radio range and permit an
analysis of the relation of the burst's prompt emission to
its afterglow in the context of the `fireball' model.

\section{OBSERVATIONS}

\subsection{Prompt $\gamma$-ray emission}

In Fig. 1 we show the BATSE light curves of \g123\ in two energy
bands.  The burst profile is dominated by two peaks, each lasting
about 8 seconds (FWHM), separated by 12 seconds, followed by a
shoulder lasting some 40 seconds.  The low-energy emission persists
longer than the high-energy emission, i.e., GRB\,990123 presents a
typical case of ``hard-to-soft'' spectral evolution$^{13}$.
Particularly striking is the paucity of $>$ 300 keV emission during
the shoulder. The $T_{90}$ duration of the burst$^{14}$, as measured
in the 50 - 300 keV range is $63.30\pm 0.26$ seconds.

Using the BATSE data, we find for the peak flux (50--300 keV) and
total fluence ($>$ 20 keV) values of $F_{\rm max} = 3.8 \times
10^{-6}$ erg\,cm$^{-2}$\,s$^{-1}$, and $E_{\rm b} = 3.0 \times
10^{-4}$ erg\,cm$^{-2}$.  The redshift ($z=1.61$) of the optical
afterglow, obtained from Fe and Mg absorption lines$^{15}$ implies a
minimum luminosity distance of 11 Gpc, and a total rest-frame ($>$ 20
keV) $\gamma$-ray energy of $4 \times 10^{54}$ erg, using $H_0 = 70$
km\,s$^{-1}$\,Mpc$^{-1}$, and $\Omega_0 = 0.3$ and assuming isotropic
emission.  To investigate the relation between the optical emission
detected with ROTSE and the $\gamma$-ray burst we made spectral fits
to the $\gamma$-ray data during the time intervals corresponding to
the first three ROTSE images.  Fig. 2 compares our gamma-ray fits to
the simultaneous optical observations by ROTSE.  Extending the
low-energy power-law to optical frequencies, even allowing for the
slope uncertainties, we find that in all three spectra the optical
points fall well above the low-energy extension of the $\gamma$-ray
spectrum, independent of the slope of the latter.

\subsection{Multi-wavelength afterglow emission}

We observed the afterglow of \g123 in optical, infra-red, sub-mm, mm
and radio passbands (see Tables 1, 2 and 3).  The optical R-band light
curve after the first 0.1 day is well described by a power law decay,
$F_\nu = F_0 \cdot t^{\alpha}$ ($t$ is the time since the burst): we
find $\alpha = -1.12 \pm 0.03$ ($\chi^{2}_{\rm r} = 30/28$).  To
account for the presence of light from the underlying host
galaxy$^{16}$ we have also fit a model consisting of a power-law decay
plus a constant flux; we find $\alpha= -1.14 \pm 0.03$ and $R_{\rm
host}> 24.9$ ($\chi^{2}_{\rm r} = 30/27$).  The infra-red observations
show a decay that is consistent with that in the R-band; $\alpha_{\rm
H} = -0.94 \pm 0.22$ and (from Ref. 17) $\alpha_{\rm K} = -1.14 \pm
0.08$.  The late-time power law decay connects smoothly with the last
three data points of the ROTSE observations (Fig. 3), but not
with the first three: the slope of the light curve before these three
points is much steeper.

We detect the radio afterglow at 4.88 GHz at the improved optical
position$^{18}$, using the Westerbork radio telescope. An r.m.s.-noise
weighted flux density average gives $F_{\rm 4.88 GHz} = 118 \pm 40
\mu$Jy (average epoch January 24.46 UT). After January 26 we do not
detect the source at 4.88 GHz ($F_{\rm 4.88 GHz} = -7 \pm 31 \mu$Jy;
weighted average of the January 26, 28 and 29 data).  Similar behavior
was observed$^{19}$ at 8.46 GHz: initially the source was not detected
($<$68
$\mu$Jy; $2\sigma$; January 23.63 UT), then it was detected (260 $\pm$
32 $\mu$Jy; January 24.65 UT) and not detected after January
26. Such radio behavior is unique, both for its early
appearance as well as its rapid decline.  The source was not detected
at 1.38, 15, 85, 140, 225, 350 and 670 GHz (see Table 2).

We have reconstructed the radio to X-ray afterglow spectrum on January
24.65 UT (see Fig. 4). This interval was selected for the long
wavelength coverage possible at the only time of radio detections.
Describing the spectrum by a power law ($F_{\nu} \propto \nu^{\beta}$)
we find that in the optical range $\beta = -0.75 \pm 0.23$, 
between the optical and X-ray wavebands the spectral slope
is $\beta = -0.67 \pm 0.02$, while in the radio range the spectrum
rises as $\beta = +1.4 \pm 0.7$.

\section{DISCUSSION}

Among the six GRBs with known redshifts one other (GRB\,980329) has a
total energy (assuming isotropic emission) in excess of $10^{54}$
erg$^{20}$. Of course, beaming of the $\gamma$-ray emission (i.e.,
strongly anisotropic emission) may make the total energy substantially
smaller, and until we have a good understanding of beaming the total
energy is not a severe constraint on theoretical models of the initial
explosion.

If one assumes that the ROTSE emission comes from a non-relativistic
source of size $ct$, then we can obtain a lower limit to the
brightness temperature since the restframe $T_b$ cannot exceed the
Compton limit of $10^{12}$\,K. The observed $T_b\gax 10^{17}$\,K, is a
factor $\Gamma^3$ times the restframe value, which  means that the
bulk Lorentz factor, $\Gamma \gax 50$, confirming the highly
relativistic nature of the GRB source. In view of this, and the
independent evidence$^{2,21,22}$ that GRBs and their afterglows are
emitted by relativistic shocks, we will discuss our results in terms
of this `fireball' model.

The ROTSE observations$^{12}$ have shown that the prompt optical and
$\gamma$-ray light curves do not track each other. We show that the
prompt optical emission detected with ROTSE during GRB~990123 is not a
simple extrapolation of the $\gamma$-ray burst to much lower energies,
but that there is a smooth connection between the optical emission
measured 4.5 minutes after the burst and the afterglow detected later.

The reverse shock could cause emission$^{3,4}$ that peaks in the
optical waveband and is observed only during or just after the
$\gamma$-ray burst, because the shock takes of order the duration of
the burst to travel through the ejecta and then stops emitting
quickly. The observed properties of GRB~990123 appear to fit this
model quite well. If this interpretation is correct, GRB~990123 would
be the first burst in which all three emitting regions have been seen:
internal shocks causing the $\gamma$-ray burst, the reverse shock
causing the prompt optical flash, and the forward shock causing the
afterglow. The relatively smooth connection between the prompt optical
emission and the afterglow fits the expectation that the reverse and
forward shock start at about the same time. This model requires
roughly equipartition magnetic fields in the ejecta, as does the
prompt $\gamma$-ray emission.

The radio emission of GRB\,990123 is unique both due to its very early
appearance and its rapid decline. A look at the broad-band spectrum on
Jan 24.65 (Fig. 4) reveals why this was so: it is a power law
connecting the X-ray to optical wavebands, which peaks in the several
tens of GHz range and turns over towards even lower frequencies.  The
synchrotron peak frequency does not fall very much above, or is even
below, the radio waveband. This location is very different from that
of GRB\,970508, for which the peak was still in the mm region after 12
days$^{2}$, and that of GRB\,971214, for which the peak was in the
optical/near infra-red waveband after 0.6 days$^{23}$. We infer from
this that the synchrotron peak frequency may span a large range in
$\gamma$-ray burst afterglows. The rapid decline of the radio flux,
caused by the low value of $\nu_{\rm m}$, may explain why some
$\gamma$-ray bursts are not detected at radio wavelengths.

Regardless of the location of the peak, the steeply rising radio
spectrum implies that the self-absorption frequency, $\nu_{\rm a}$
$\sim$ 30\,GHz, similar to earlier bursts.  In relativistic blast wave
models$^{24}$ the flux at the peak is constant during the blast wave
evolution, although the peak frequency decreases rapidly, as
$t^{-3/2}$. In Fig 4. two possible low frequency extrapolations are
shown. Both have a self-absorption frequency, $\nu_{\rm a}$, close to
the radio, but the lower one still has a synchrotron peak frequency,
$\nu_{\rm m}>\nu_{\rm a}\sim30$\,GHz, so that there is an optically
thin synchrotron regime, where the spectrum follows the standard
low-frequency power law $F_{\nu} \propto \nu^{1/3}$. This implies that
the radio flux should still rise after Jan 24.65, clearly inconsistent
with the non-detections a few days after Jan 24.65. The second
possibility is that $\nu_{\rm m}<\nu_{\rm a}$ already on Jan. 24.65;
then the flux at frequencies at and below $\nu_{\rm a}$ would follow
the optical decline if $\nu_{\rm a}$ were constant. However, $\nu_{\rm
a}$ now begins to fall somewhat as well, and this will lessen the rate
of decline at frequencies below $\nu_{\rm a}$ (very much below
$\nu_{\rm a}$ it even rises). Here we estimate that on Jan 24.65 the
synchrotron peak frequency was at about 30\,GHz, as was $\nu_{\rm a}$,
causing a decline of the VLA and WSRT fluxes approximately as 
$t^{-0.9}$.  The first non-detections by the VLA$^{19}$ and WSRT after
their respective detections are consistent with that decay rate.

The peak flux on Jan. 24.65 (Fig. 4) is less than 2 mJy, while the
last three ROTSE exposures have fluxes of $\sim$ 10 mJy. If these
exposures captured the early afterglow, the peak flux apparently
decreased between 7.5 minutes and $\sim$ 1.2 days after the burst,
contrary to the prediction of a constant peak flux during the blast
wave evolution. There are, however, several possible explanations for
the apparent decrease of the peak flux. If the peak frequency is less
than the self-absorption frequency $\nu_{\rm a}$, then the peak flux
will be lowered relative to its unabsorbed value. Or part of the ROTSE
flux could be due to a decaying prompt optical component and thus the
afterglow component at $t=7.5$ min. could be less than 10 mJy.
Finally, the afterglow peak flux may truly decay with time. The data,
therefore, do not allow a strong distinction between a constant and a
decreasing peak flux of the afterglow spectrum.

A comparison of GRB\,990123 with other well-studied GRBs suggests some
patterns. The cooling frequency of GRB\,970508, which separates
quickly cooling synchrotron electrons from slowly cooling ones, was in
the optical/near infra-red after 1.5 days$^{2}$. For GRB\,990123 and
all other bursts in which the optical-to-X-ray afterglow spectrum
could be reconstructed for the few days after the burst, any cooling
break was located at or above X-ray frequencies. Both a higher cooling
frequency and a lower peak frequency can be explained by a difference
in one shock parameter: the magnetic field in the forward shock
region. The lower field also causes a lower peak flux$^{24,25}$, which
with the low peak frequency conspires to make the radio emission weak
and brief. The field energy density for the radio-quiet GRB\,971214 is
estimated to be less than $10^{-5}$ times the equipartition value, and
more than 1000 times weaker than in GRB\,970508$^{25}$; the value for
GRB\,990123 may be as low as $10^{-6}$ times equipartition. This
suggests we should distinguish between low-field afterglows, which are
short and dim in radio, and high-field afterglows, which are bright
and long-lived in radio. The absence of detected radio emission in
previous bursts with afterglows can then be explained by assuming they
were low-field afterglows. We conjecture that such differences in
field strength reflect differences in energy flow from the central
engine, e.g., in the form of a Poynting wind.

The United Kingdom Infra-Red Telescope is operated by the Joint
Astronomy Centre on behalf of the U.K. Particle Physics and Astronomy
Research Council.  Observations at the Wise Observatory are supported
by the Basic Science Foundation of the Israeli Academy of Sciences.
The JCMT is operated by the Joint Astronomy Centre on behalf of the UK
Particle Physics and Astronomy Research Council, the Netherlands
Organisation for Pure Research, and the National Research Council of
Canada.  We thank G. Watt, I. Robson, and J. van der Hulst for
authorizing the JCMT observations.  We wish to thank Peter Meikle,
Paul Smith, Di Harmer and the KPNO GRB team for their observations,
and George Jacoby for providing BVRI photometric standards. TJG is
supported by NFRA, PMV by the NWO Spinoza grant. CK, DLB, KH and MSB
are supported by NASA.

\bigskip
\centerline{\bf References}

1. Piran, T. Gamma-Ray Bursts and the Fireball Model. {\it Physics
Report} (in the press); preprint http://xxx.lanl.gov,
astro-ph/9810256 

2. Galama, T.J. {\it et al.} The radio-to-X-ray spectrum of GRB 970508
on 1997 May 21.0 UT. {\it Astrophys. J.} {\bf 500}, L97-L101
(1998). 

3. M\'esz\'aros, P. and Rees, M.J. Optical and long-wavelength
afterglow from gamma-ray bursts. {\it Astrophys. J.} {\bf 476},
232-237 (1997). 

4. Sari, R. \& Piran, T. Predictions for the very early afterglow and
the
optical flash. preprint http://xxx.lanl.gov, astro-ph/9901338
(1998).

5. Katz, J.I.  Low-frequency spectra of gamma-ray bursts. {\it
Astrophys. J.} {\bf 432}, L107-L109 (1994). 

6. Costa, E. {\it et al.} Discovery of an X-ray afterglow associated
with
the $\gamma$-ray burst of 28 February 1997. {\it Nature} {\bf 387},
783-785 (1997). 

7. Feroci, M., {\it et al.} BeppoSAX follow-up search for the X-ray
afterglow of GRB\,970111. {\it Astron. Astrophys.} {\bf 332}, L29-L32
(1998). 

8. in 't Zand, J.J.M, {\it et al.} Gamma-Ray Burst 980329 and its
X-ray afterglow. {\it Astrophys. J.} {\bf 505}, L119-L122 (1998).

9. Nicastro, L., {\it et al.} BeppoSAX observations of
GRB\,970402. {\it 
Astron. Astrophys.} {\bf 338}, L17-L20 (1998). 

10. Piro, L., {\it et al.} Evidence for a late-time outburst of the
X-ray
afterglow of GB970508 from BeppoSAX {\it 
Astron. Astrophys.} {\bf 331}, L41-L44 (1998). 

11. Connors, A. \& Hueter, G.J. The X-ray characteristics of a
classical
gamma-ray burst and its afterglow. {\it Astrophys. J.}  {\bf 501},
307-324 (1998). 

12. Akerlof, C.W. and McKay, T.A. {\it Nature} (in the press)
(1999).

13. Ford, L.A. {\it et al.} BATSE observations of gamma-ray burst
spectra. 2: Peak energy evolution in bright, long bursts. {\it
Astrophys. J.} {\bf 439}, 307-321 (1995). 

14. Kouveliotou, C. {\it et al.} Identification of two classes of
gamma-ray bursts. {\it Astrophys. J.} {\bf 413}, L101-L104 (1993). 

15. Kelson, D.D. {\it et al.} {\it IAU Circ.} No. 7096 (1999). 

16. Fruchter, A. {\it et al.} {\it GCN Circ.} No. 255 (1999). 

17. Bloom, J.S. {\it et al.} {\it GCN Circ.} No. 240 (1999). 

18. Bloom, J.S.,  Gal, R.R., Lubin, L.L. Mulchaey, J., Odewahn, S.C.,
Kulkarni, S.R. {\it GCN Circ.} No. 206 (1999).

19. Kulkarni, S.R., Frail, D.A. {\it GCN Circ.} No. 239 (1999). 

20. Fruchter, A. Was GRB\,980329 at z$\sim$5? {\it Astrophys. J.} (in
the press); preprint http://xxx.lanl.gov, astro-ph/9810224 (1998).

21. Waxman, E. Angular size and emission timescales of relativistic
fireballs. {\it Astrophys. J.} {\bf 491}, L19--L22 (1997)

22. Wijers, R. A. M.~J., Rees, M.~J., and {M\'esz\'aros}, P.
Shocked by GRB 970228: the afterglow of a cosmological fireball
{\it Mon. Not. R. Astron. Soc.} {\bf 288}, L51--L56 (1997)

23. Ramaprakash, A.N. {\it et al.} The energetic afterglow of the
$\gamma$-ray burst of 14 December 1997. {\it Nature} {\bf 393},
43-46 (1998). 

24. Sari, R., Piran, T. \& Narayan, R. Spectra and light curves of
gamma-ray burst afterglows. {\it Astrophys. J.} {\bf 497}, L17-L20
(1998). 

25. Wijers, R.A.M.J. \& Galama, T.J. Physical parameters of GRB 970508
and
GRB 971214 from their afterglow synchrotron 
emission. {\it Astrophys. J.} (in the press);
preprint http://xxx.lanl.gov, astro-ph/9805341 (1998).

26. Band, D. {\it et al.} BATSE observations of gamma-ray burst
spectra. I
- Spectral diversity. {\it Astrophys. J.} {\bf 413}, 281-292
(1993). 

27. Gal, R.R, Odewahn, S.C., Bloom, S.J., Kulkarni, S.R., Frail,  
D.A. {\it GCN Circ.} No. 207 (1999)

28. Garnavich, P., Jha, S., Stanek, K., Garcia, M. {\it GCN Circ.}
No. 215
(1999) 

29. Masetti, N., Palazzi, E., Pian, E., Frontera, F., Barolini, C.,
Guarnieri, A., P iccioni, A., Valentini, G., Costa, E. {\it GCN Circ.}
No. 233 (1999)

30. Sokolov, V., Zharikov, S., Nicastro,
L., Feroci, M., Palazzi, E. {\it GCN Circ.} No. 209 (1999).

31. Zhu, J., Chen, J.S. and Zhang, H.T. {\it GCN Circ.} No. 217
(1999).

32. Yadigaroglu, I.A, Halpern, J.P., Uglesich, R., Kemp, J. {\it GCN
Circ.} No. 242 (1999). 

33. Yadigaroglu, I.A, Halpern, J.P., Uglesich, R., Kemp, J. {\it GCN
Circ.} No. 248 (1999). 

34. Frail, D.A., Kulkarni, S.R., {\it GCN Circ.}, No. 211 (1999). 

35. Heise, J. {\it et al.} {\it IAU Circ.} No. 7099 (1999).  

36. Bessell, M.S. UBVRI photometry. II - The Cousins VRI system, its
temperature and absolute flux calibration, and relevance for
two-dimensional photometry. {\it Publ. Astron. Soc. Pacif.}  {\bf
91}, 589-607 (1979). 

37. Bessell, M.S. and Brett, J.M. JHKLM photometry - Standard systems,
passbands, and intrinsic colors. {\it Publ. Astron. Soc. Pacif.}  {\bf
100}, 1134-1151 (1988). 

38. Schlegel, D.J., Finkbeiner, D.P. and Davis, M. Maps of Dust
Infrared
Emission for Use in Estimation of Reddening and Cosmic Microwave
Background Radiation Foregrounds. {\it Astrophys. J.} {\bf 500},
525-553
(1998). 

39. Granot, J., Piran, T. and Sari, R. Synchrotron Self Absorption in
GRB Afterglow {\it Astrophys. J.} (in the press); preprint
http://xxx.lanl.gov, astro-ph/9808007 (1999).

40. Baars, J.W.M., Genzel, R., Pauliny-Toth, I.I.K., Witzel,
A. The absolute spectrum of Cas A - An accurate flux density scale and
a set of secondary calibrators. {\it Astronomy \& Astrophysics} {\bf
61}, 99-106 (1977).

41. Guilloteau S. {\it et al.} The IRAM interferometer on
Plateau de Bure. {\it Astronomy \& Astrophysics} {\bf 262}, 624-633
(1992).

42. Baars, J.W.M., Hooghoudt, B.G., Mezger, P.G., De Jonge,
M.J., The IRAM 30-m millimeter radio telescope on Pico Veleta, Spain.
{\it Astronomy \& Astrophysics} {\bf 175}, 319-326 (1987).

43. Reuter H.-P. and Kramer C. The mm-to-submm continuum
spectra of W 3(OH) and K 3-50A. {\it Astronomy \& Astrophysics} {\bf
339}, 183-186 (1998).

44. Greve, A., Neri, R. and Sievers, A. The gain-elevation
correction of the IRAM 30-m telescope. {\it Astronomy \& Astrophysics
Suppl.} {\bf 132}, 413-416 (1998).

45. Holland, W. S., et al. SCUBA: A common-user
submillimetre camera operating on the James Clerk Maxwell
Telescope. {\it Mon. Not. R. Astron. Soc.} (in the
press); preprint http://xxx.lanl.gov, astro-ph/9809122
(1999). 

46. Smith, I. A., et al. SCUBA sub-millimeter observations
of gamma-ray bursters I. GRB 970508, 971214, 980326, 980329, 980519,
980703.  {\it Astronomy \& Astrophysics} (in the
press); preprint http://xxx.lanl.gov, astro-ph/9811026
(1999). 

47. Landolt, A.U., UBVRI photometric standard stars in the magnitude 
range 11.5 $<V<$16.0 around the celestial equator. {\it Astron. J.}
{\bf 104}, 340-376 (1992).

48. Casali, M.M. and Hawarden, T.G. JCMT-UKIRT Newsletter, No. 3, 33
(1992) 

\newpage

\begin{figure}
\centerline{\psfig{figure=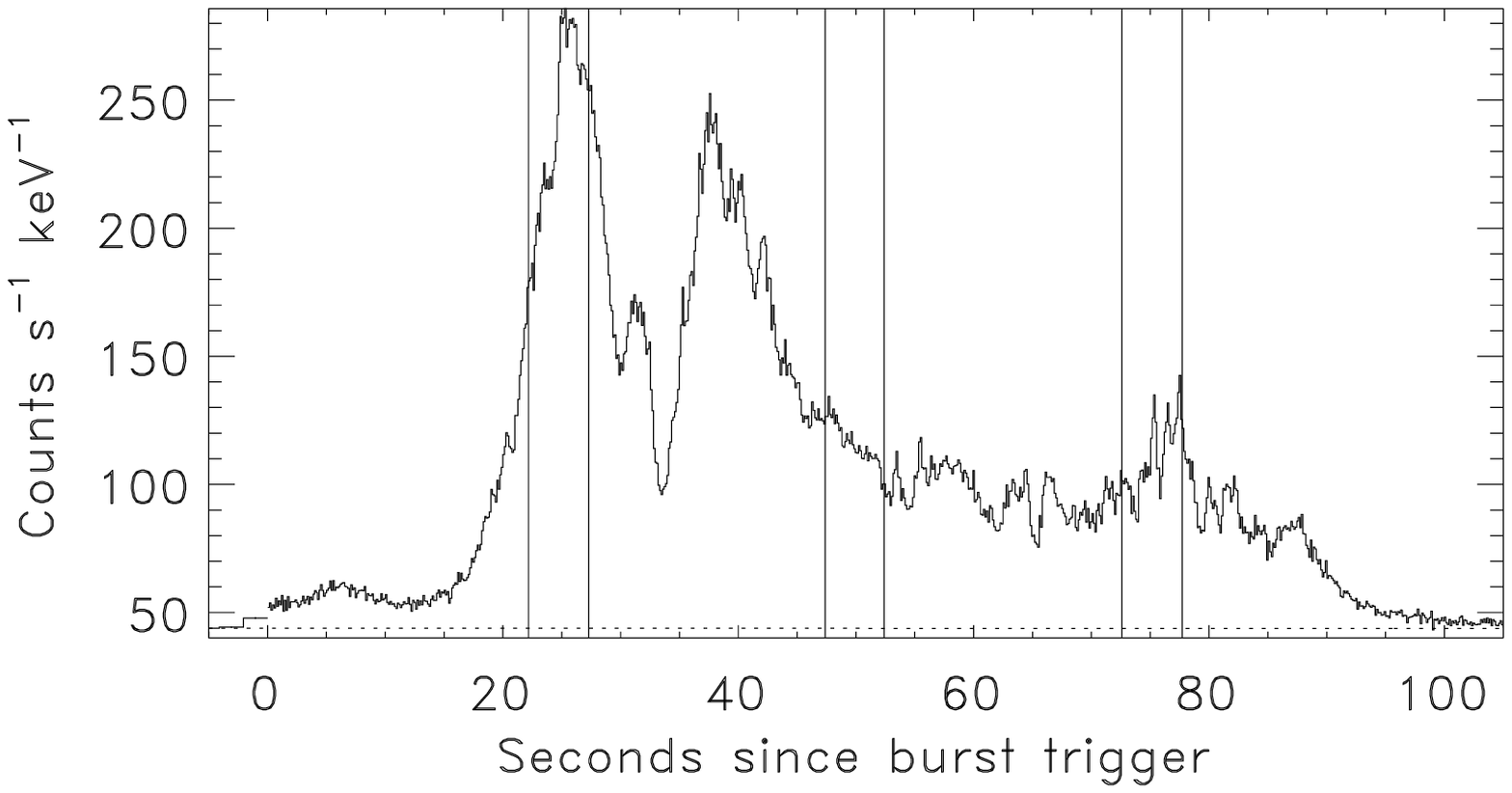,width=12cm}}
\centerline{\psfig{figure=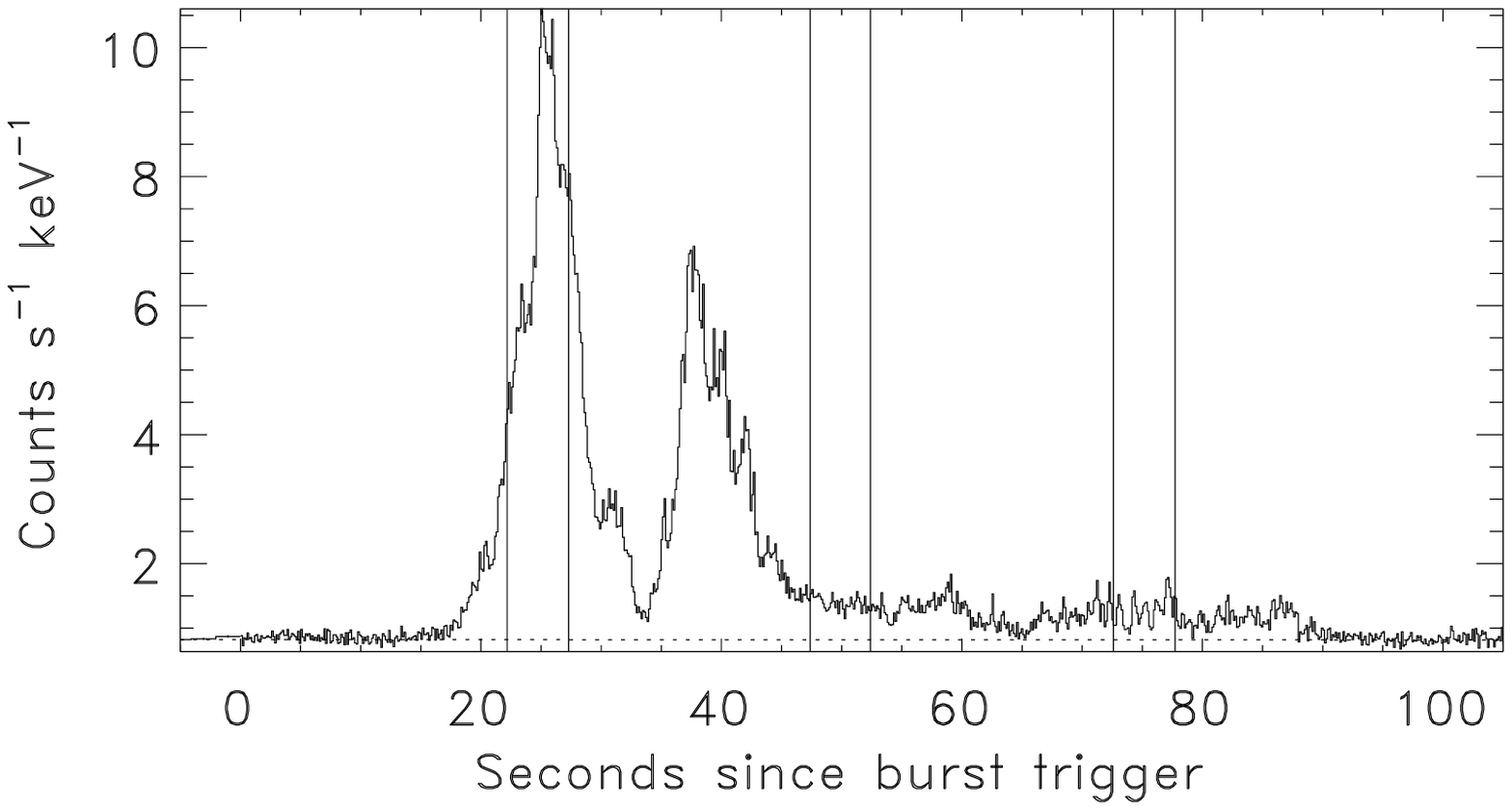,width=12cm}}
\caption{The two panels show the BATSE light curves of \g123\ in two
energy ranges, 25--230 keV (top) and 320--1800 keV (bottom).  Times
are relative to the BATSE trigger time of UT 9h 46m 56.1s on January
23.  The vertical lines indicate the time intervals of the first three
ROTSE observations$^{12}$.}
\end{figure}

\begin{figure}
\centerline{\psfig{figure=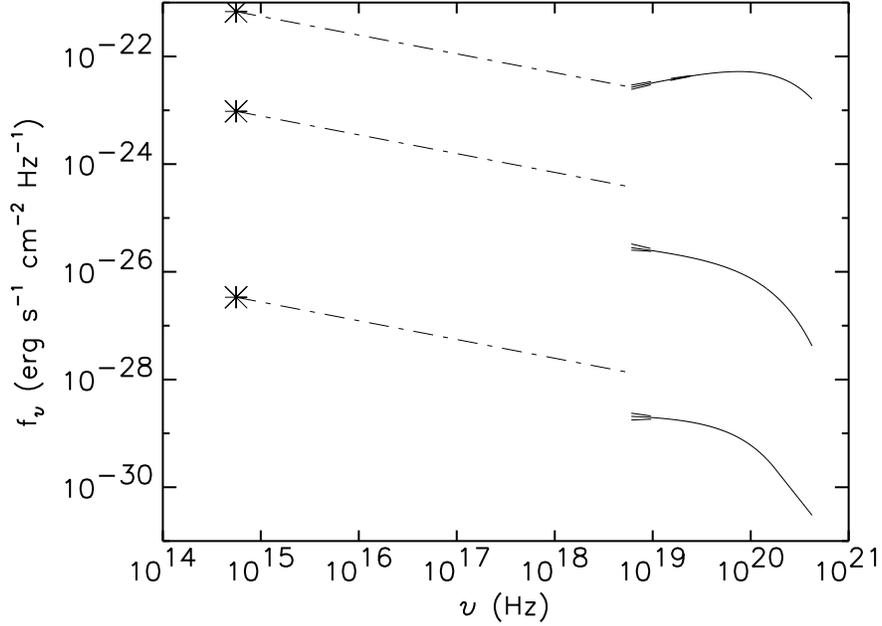,width=12cm}}
\caption{
Optical flux and gamma-ray spectra for the time intervals of the 3
ROTSE observations. The gamma-ray curves (solid) are fits to BATSE
spectra using the Band function$^{26}$, which consists of a low-energy
power law with an exponential cutoff which merges smoothly with a
high-energy power law. The curves for the first (third) time interval
have been shifted up (down) by a factor of 1000.  The data point on the
left is the reported ROTSE magnitude converted into a flux density in
the middle of the V band. The dashed curves on the low energy end of
the gamma-ray spectra show the $3\sigma$ variations in the low energy
spectral index. The dot-dashed curves are the extrapolation of the
optical flux to the X-ray band using the power law which connects the
observed optical and lowest-energy $\gamma$-ray fluxes for the first
ROTSE time interval.}
\end{figure}

\begin{figure}
\centerline{\psfig{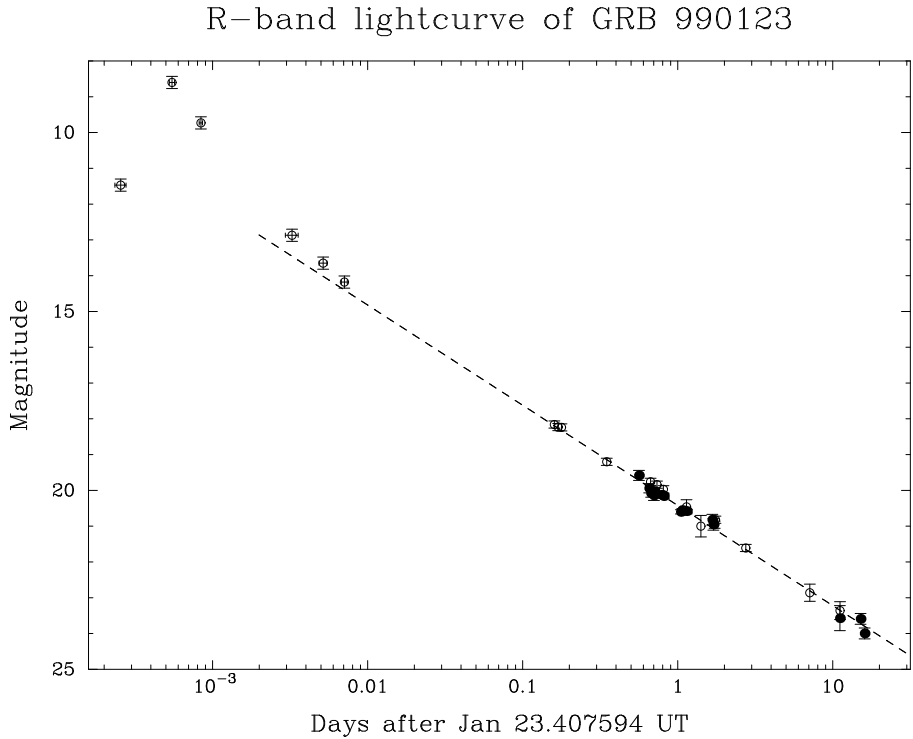}}
\caption{
R-band light curve of the afterglow of \g123. The filled circles
indicate results of our observations. The open circles are data taken
from Refs 27-33: we determined the corresponding magnitude for our
calibration. The dashed line indicates a power law fit to the light
curve (for $t > 0.1$ days), which has exponent $-1.12 \pm 0.03$. Also
included are the ROTSE data$^{12}$. ROTSE uses an unfiltered CCD, but
an equivalent V band magnitude is reported; here we adopted a 0.1m
error for the ROTSE data points and assuming a constant color we have
applied a color correction of V-R = 0.35 $\pm$ 0.14. The power law fit
is extrapolated backward; it gives a splendid description of the last
three ROTSE data points.}
\end{figure}

\begin{figure}
\centerline{\psfig{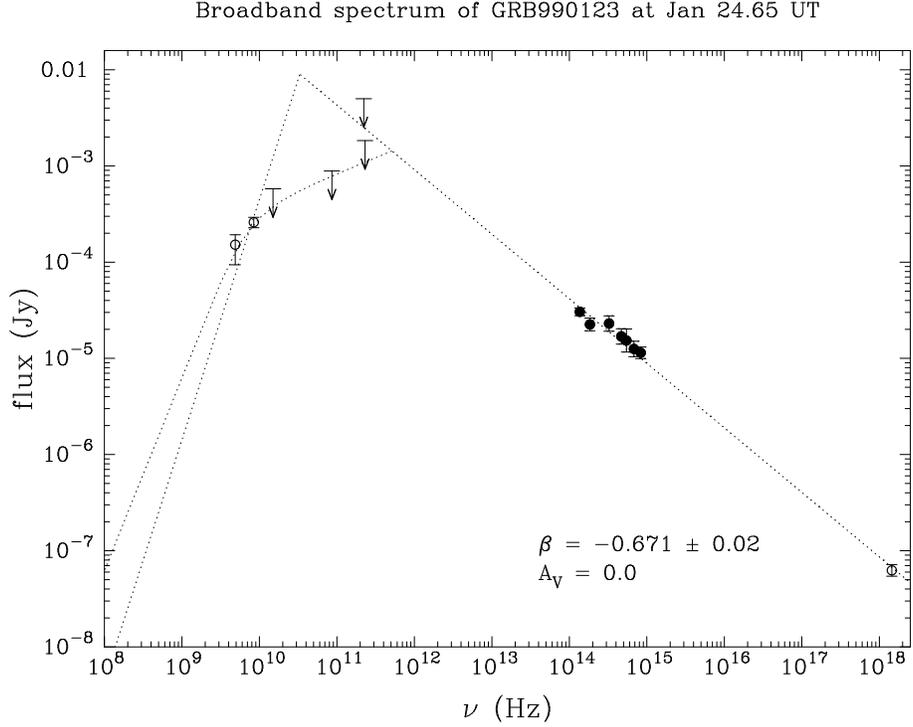}}
\caption{
The spectral flux distribution of the afterglow at 1999 January 24.65
UT.  Shown are the WSRT 4.88 GHz detection (this paper), the VLA 8.46
GHz detection$^{34}$, the KPNO 4-m U band, WIYN B,V,R and I and UKIRT
H observations (this paper), the K band detection$^{17}$, and the
X-ray (2-10 keV) flux$^{35}$. Also shown are the Ryle 15 GHz, PdBI 86
and 230 GHz and the JCMT 220 GHz 2-$\sigma$ upper limits (this paper;
we scaled these values back to the central time of the epoch by assuming
that the source flux exhibits a $t^{-1.1}$ decay, i.e., we used
conservative limits).  The photometric calibration has been taken from
Ref. 36 for B, V, R and I and Ref. 37 for H.  We corrected the optical
fluxes for Galactic foreground absorption (A$_V$ = 0.053, as inferred
from the dust maps of Ref. 38). If more than one value per filter was
available around the central time of the epoch, we took their weighted
average. All values were brought to the same epoch by applying a
correction using the slope of the fitted light curve. We fitted the
resulting optical to X-ray spectral flux distribution with a power law
and an exponential optical extinction law, $F_\nu \propto
\nu^{\beta}e^{-\tau}$, where we assume that the extinction optical
depth, $\tau \propto \nu$. The fit provides a negative extinction
(A$_V < 0.19$; 90 \% confidence) and was subsequently fixed at
zero. We find $\beta = -0.67 \pm 0.02$ ($A_V = 0$; $\chi^{2}_{\rm r}$
= 6/7); indicated is the resulting fit. Also shown for reference are
two possible extrapolations of the low-frequency part of the spectrum:
(i) a self-absorbed spectrum $F_{\nu} = F_{\nu_{\rm a}}(\nu/\nu_{\rm
a})^{2}(1-\exp[-(\nu/\nu_{\rm a})^{-5/3}])$ (Ref 39), for a
self-absorption frequency, $\nu_{\rm a}$ = 7.5 GHz and a flux density,
$F_{\nu_{\rm a}}$ = 350 $\mu$Jy, and (ii) the classic spectrum of a
self-absorbed radio source with asymptotic spectral index +5/2. Note
that realistic spectra are much rounder at the peak than the simple
broken power law spectra shown (see the discussion for the details).}
\end{figure}



\scriptsize

\def\arcsec{\hbox{$^{\prime\prime}$}}
\begin{table}
\begin{center}
\begin{tabular}{lccrc}
\multicolumn{5}{c}{\textbf{Optical and Infra-red observations of
GRB~990123}} \\
\hline
UT date &
magnitude &
filter &
exp. time &
telescope/reference \\
(1999 Jan) &
&
&
(seconds) &
\\
\hline
Jan 23.970 &     19.58 $\pm$  0.14 &   R  & 3600 & UPSO 1-m\\
Jan 24.028 &     20.42 $\pm$  0.10 &   B  & 3000 & UPSO 1-m\\
Jan 24.999 &     21.82 $\pm$  0.12 &   B  & 3600 & UPSO 1-m\\
Jan 25.029 &     21.14 $\pm$  0.16 &   V  &  600 & UPSO 1-m\\
Jan 24.066 &     20.38 $\pm$  0.26 &   B  &  120 & Wise 1-m\\
Jan 24.062 &     19.94 $\pm$  0.13 &   R  &  600 & Wise 1-m\\
Jan 24.081 &     20.08 $\pm$  0.11 &   R  &  600 & Wise 1-m\\
Jan 24.110 &     20.14 $\pm$  0.14 &   R  &  600 & Wise 1-m\\
Jan 24.116 &     20.02 $\pm$  0.09 &   R  &  600 & Wise 1-m\\
Jan 25.079 &     20.82 $\pm$  0.15 &   R  &  900 & Wise 1-m\\
Jan 25.092 &     21.75 $\pm$  0.21 &   B  & 1200 & Wise 1-m\\
Jan 25.105 &     20.95 $\pm$  0.16 &   R  &  900 & Wise 1-m\\
Jan 24.198 &     20.12 $\pm$  0.06 &   R  & 1000 & JKT\\
Jan 24.210 &     20.14 $\pm$  0.05 &   R  & 1000 & JKT\\
Jan 24.222 &     20.16 $\pm$  0.06 &   R  & 1000 & JKT\\
Jan 24.456 &     20.60 $\pm$  0.06 &   R  &  900 & WIYN\\
Jan 24.477 &     20.55 $\pm$  0.04 &   R  &  600 & WIYN\\
Jan 24.487 &     20.14 $\pm$  0.08 &   I  &  600 & WIYN\\
Jan 24.496 &     20.89 $\pm$  0.06 &   V  &  600 & WIYN\\
Jan 24.505 &     21.21 $\pm$  0.04 &   B  &  600 & WIYN\\
Jan 24.515 &     20.57 $\pm$  0.04 &   R  &  600 & WIYN\\
Jan 24.544 &     20.96 $\pm$  0.08 &   V  &  600 & WIYN\\
Jan 24.553 &     20.58 $\pm$  0.05 &   R  &  600 & WIYN\\
Feb  3.519 &	 23.57 $\pm$  0.35 &   R  &  900 & WIYN\\
Feb  3.533 &	$>$ 22.8      	   &   I  &  900 & WIYN\\
Feb  7.513 &     23.59  $\pm$ 0.15 &   R  & 3500 & WIYN\\
Feb  8.460 &     24.00  $\pm$ 0.15 &   R  & 3600 & WIYN\\
Jan 24.497 &     20.34 $\pm$  0.10 &   U  &  900 & KPNO 4-m\\
Jan 24.660 &     19.14 $\pm$  0.13 &   H  & 1020 & UKIRT\\  
Jan 27.660 &     20.39 $\pm$  0.26 &   H  & 1080 & UKIRT\\
\end{tabular}
\caption[]{\scriptsize The log of the optical and infra-red observations.  
Eight reference stars (see Table 3) were used to obtain the
differential magnitude of the optical transient (OT) in each
observation. Telescopes: WIYN 3.5-m telescope at Kitt Peak (WIYN),
Uttar Pradesh State Observatory 1-m telescope (UPSO 1-m), Wise
Observatory 1-m telescope (Wise 1-m), Jakobus Kapteyn Telescope, La
Palma (JKT), Kitt Peak National Observatory 4-m Mayall telescope (KPNO
4-m), United Kingdom Infra-Red Telescope, Hawaii (UKIRT).  Instruments
and CCDs used: UPSO 1-m: CCD Camera, 512 $\times$ 512 CCD, 0.76
\arcsec /pixel; Wise 1-m: TEK 1k $\times$ 1k CCD, 0.70\arcsec /pixel;
JKT: TEK 2k $\times$ 2k CCD, 0.33\arcsec/pixel; WIYN: TEK 2k $\times$
2k CCD, 0.197\arcsec /pixel; KPNO 4-m: 8k $\times$ 8k CCD Mosaic
Camera, 0.258\arcsec/pixel; UKIRT: IRCAM3 with FPA42 256 $\times$ 256
detector, 0.29\arcsec /pixel.}
\end{center}
\end{table}

\def\degr{\hbox{$^\circ$}}
\def\arcmin{\hbox{$^\prime$}}
\def\arcsec{\hbox{$^{\prime\prime}$}}
\def\fd{\hbox{$.\!\!^{\rm d}$}}
\def\fh{\hbox{$.\!\!^{\rm h}$}}
\def\fm{\hbox{$.\!\!^{\rm m}$}}
\def\fs{\hbox{$.\!\!^{\rm s}$}}
\def\fdg{\hbox{$.\!\!^\circ$}}
\def\farcm{\hbox{$.\mkern-4mu^\prime$}}
\def\farcs{\hbox{$.\!\!^{\prime\prime}$}}         
\def\gsim{\mathrel{\hbox{\rlap{\lower.55ex \hbox {$\sim$}}
                   \kern-.3em \raise.4ex \hbox{$>$}}}}
\def\lsim{\mathrel{\hbox{\rlap{\lower.55ex \hbox {$\sim$}}
                   \kern-.3em \raise.4ex \hbox{$<$}}}}
\def\arcsec{\hbox{$^{\prime\prime}$}}

\tiny
\begin{table}[!ht]
\begin{center}
\begin{tabular}{clcccc}
\multicolumn{5}{c}{\textbf{Radio and (sub) mm observations of
GRB~990123}} \\
\hline
Freq   & UT date 1999  & duration   &     flux   & Reference\\
      & (central) & (hours) & microJansky \\            
\hline
1.380 & Jan 27.20  & 8.4   & 37  $\pm$ 22   & WSRT \\
4.88 & Jan 24.28  & 12    & 104  $\pm$ 42  & WSRT \\
4.88 & Jan 25.47  & 2.9   & 164  $\pm$ 100 & WSRT \\
4.88 & Jan 26.29  & 1.4   & --65  $\pm$ 130 & WSRT  \\
4.88 & Jan 28.27  & 12    & --3   $\pm$ 41  & WSRT \\
4.88 & Jan 29.17  & 7.4   & --9.3 $\pm$ 55  & WSRT \\
4.88 & Jan 30.27  & 12    & --74  $\pm$ 44   & WSRT \\
4.88 & Jan 31.26  & 12    &  20 $\pm$  41      & WSRT \\
4.88 & Feb 6.25 & 12  &  30   $\pm$ 44   & WSRT \\
4.88 & Feb 7.13 & 6.7   & 112 $\pm$ 85   & WSRT \\
8.46  & Jan 23.63   &       &  $<$68 ($2\sigma$) & VLA \\
8.46  & Jan 24.65   &       & 260 $\pm$ 32   & VLA \\
8.46  & Jan 26      &       & $<$78 ($2\sigma$) & VLA \\
8.46  & Jan 27      &       & $<$50 ($2\sigma$) & VLA \\
8.46  & Jan 28      &       & $<$50 ($2\sigma$) & VLA \\
15.0  & Jan 25.26  & 9.9   & 160 $\pm$ 180  & Ryle \\
15.0  & Jan 26.27  & 10.4  & --12 $\pm$ 180  & Ryle \\
15.0  & Jan 29.20  & 6.4   & --197$\pm$ 200 & Ryle \\
15.0  & Jan 30.40  & 3.7   & 106 $\pm$ 420 & Ryle \\
15.0  & Jan 31.39  & 3.8   &  48 $\pm$ 300 & Ryle \\
15.0  & Feb 4.39  & 3.6   & 378 $\pm$ 270 & Ryle \\
15.0  & Feb 6.37   & 4.1   &--140 $\pm$ 300 & Ryle \\
15.0  & Feb 8.36   & 2.8   & --16 $\pm$ 280 & Ryle \\
86.27 & Jan 25.18  & 4.5   & (1.0 $\pm$ 2.9)$\times10^2$ & PdBI \\
232.0 &	Jan 25.18  & 4.5   & (--1.6 $\pm$ 1.2)$\times10^3$ & PdBI \\
86.27 & Feb 1.47   & 4.5   & (--5.0 $\pm$ 3.1)$\times10^2$ & PdBI \\
212.6 & Feb 1.47   & 4.5   & (0.4 $\pm$ 1.3)$\times10^3$ & PdBI \\
86.27 & Feb 4.06   & 3.5   & (--0.3 $\pm$ 4.8)$\times10^2$ & PdBI \\
231.5 &	Feb 4.06   & 3.5   & (--2.7 $\pm$ 4.3)$\times10^3$ & PdBI \\
86.64 & Jan 28.54   & 0.42	& (--4.1 $\pm$ 9.1)$\times10^3$ & Pico
Veleta \\
142.3 & Jan 28.54   & 0.42	& (--3.7 $\pm$ 3.3)$\times10^4$ & Pico
Veleta \\
228.9 & Jan 28.54   & 0.42	& (--0.9 $\pm$ 1.9)$\times10^4$ & Pico
Veleta \\
86.64 & Jan 30.53   & 1.42	& (8.5 $\pm$ 3.5)$\times10^3$ & Pico
Veleta \\
142.3 & Jan 30.53   & 1.42	& (1.0 $\pm$ 2.9)$\times10^4$ & Pico
Veleta \\
228.9 & Jan 30.53   & 1.42	& (1.3 $\pm$ 1.2)$\times10^4$ & Pico
Veleta \\
222   & Jan 24.68   & 0.65  & (--4.1 $\pm$ 2.5)$\times10^3$  &
JCMT\footnotemark[1]\\
222   & Jan 27.83   & 0.62  & (0.7 $\pm$ 1.9)$\times10^3$  &
JCMT\footnotemark[1]\\     
353   & Jan 27.89   & 1.25  & (--3.3 $\pm$ 1.2)$\times10^3$  &
JCMT\footnotemark[1]\\      
353   & Jan 29.88   & 1.0	& (0.8 $\pm$  1.5)$\times10^3$ &
JCMT\footnotemark[1]\\     
353   & Feb 4.83    & 0.7	& (4.9 $\pm$  1.5)$\times10^3$ &
JCMT\footnotemark[1]\\   
353   & Feb 5.85    & 1.5	& (1.2 $\pm$  1.1)$\times10^3$ &
JCMT\footnotemark[1]\\   
666   & Jan 27.88   & 1.0   & (--0.2 $\pm$  1.7)$\times10^4$ &
JCMT\footnotemark[1]$^{,}$\footnotemark[2]\\
666   & Jan 29.87   & 0.75  & (--0.8 $\pm$  2.3)$\times10^4$ &
JCMT\footnotemark[1]$^{,}$\footnotemark[2]\\
666   &  Feb 4.82  & 0.5   & (1.5 $\pm$  1.8)$\times10^4$ &
JCMT\footnotemark[1]$^{,}$\footnotemark[2]\\
\hline
\end{tabular}
\caption[]{\scriptsize Summary of the radio and (sub) mm observations of
GRB~990123. 
Westerbork Synthesis Radio Telescope (WSRT): We used the
Multi-Frequency Front Ends (MFFEs)\footnote{Information on the MFFE
can be found on http://www.nfra.nl/wsrt/capabilities.htm} at 4.88 and
1.38 GHz, and the DCB continuum correlator, providing us with 8 bands
of width 10 MHz at each frequency.  The data were calibrated using
interleaved observations of 3C~48, 3C~147 and 3C~286$^{40}$.  VLA data
are taken from Ref. 19. IRAM Plateaux de Bure Interferometer
(PdBI)$^{41}$: Observations were done in five antenna configurations,
on January 25 in 5B1 and on February 1st, in 5B2. Spectral bandwidth
of the dual frequency SIS receivers at 86.2~GHz was 560~MHz (LSB
tuning), and 310~MHz (DSB tuning) at frequencies above 210~GHz. Good
weather conditions allowed to use the atmospheric phase correction
system to improve the signal to noise ratio. Results are given for
position-fixed point source fits. Pico Veleta: Observations with the
IRAM 30-m telescope on the Pico Veleta in Spain$^{42}$ were done at
86.638, 142.33 and 228.930 GHz in wobbler switching mode with SIS
receivers connected to the 1~GHz backends. Flux calibration was
relative to W3 (OH)$^{43}$, using the gain-elevation curve from
Ref. 44. James Clerk Maxwell Telescope (JCMT): Observations were made
using SCUBA$^{45}$ in the photometry mode, as for previous
bursts$^{46}$.  At 450/850 $\mu$m, GRB~990123 was in the central pixel
of the arrays (at 1350 $\mu$m there is only one bolometer).  The
secondary was chopped between the source and sky at 7 Hz to take out
small relative DC drifts between the bolometers and remove large-scale
sky variations.  The opacity at 450 and 850 $\mu$m was measured from
skydips while using the continuously monitored 1.3 mm opacity as a
guideline of any trends between the skydips. For the absolute flux
calibration the gain was measured using Mars. The $3.3 \sigma$ result
at 353 GHz on Feb 4.76 UT is presumably a statistical fluctuation,
since the result was not confirmed the following day. \\ \\
{
1.The ``integration'' time always refers to the
  on-source plus off-source time.  An 18 sec integration thus amounts
  to 9 sec on-source observation time.}
{2. The last 450/850 measurement was excluded from the
   450 reduction because of the low elevation of the source.}}
\end{center}
\end{table}

\newpage

\renewcommand{\baselinestretch}{1}
\normalsize

\begin{sidetable}
\begin{minipage}{15cm}
\begin{tabular}{lcccccccc}
\hline
ID & $\Delta$ R.A.($\arcsec$) & $\Delta$ Decl.($\arcsec$) & U & B & V
& R & I & H \\
\hline
  1 & -34.30 &  24.33 & 15.38 $\pm$ 0.01 & 15.45 $\pm$ 0.01 &  14.88
$\pm$  0.01 &    14.57 $\pm$ 0.01 &    14.28 $\pm$   0.01 & 13.54
$\pm$ 0.01 \\
  2 & -92.37 &  53.94 & 	         & 17.84 $\pm$ 0.04 &  16.83
$\pm$  0.01 &    16.00 $\pm$ 0.01 &    15.23 $\pm$   0.02 & \\
  3 & -96.92 &  48.55 & 16.88 $\pm$ 0.02 & 16.97 $\pm$ 0.02 &  16.34
$\pm$  0.01 &    15.93 $\pm$ 0.01 &    15.57 $\pm$   0.01 & \\
  4 &  49.55 &  30.08 &                  &	 &  20.34 $\pm$  0.22
&    19.66 $\pm$ 0.37 &    18.07 $\pm$   0.40 & \\
  5 & -15.19 & -29.00 &                  &	 &  20.76 $\pm$  0.32
&    19.76 $\pm$ 0.51 &    19.09 $\pm$   0.55 & 17.37 $\pm$ 0.03\\
  6 &  23.94 & -89.03 & 		 & 19.80 $\pm$ 0.23 &  18.95
$\pm$  0.06 &    18.56 $\pm$ 0.10 &    18.23 $\pm$   0.13 & \\
  7 & -30.64 & -74.52 &                  &	 &  19.86 $\pm$  0.15
&    20.08 $\pm$ 0.32 &    20.09 $\pm$   0.46 & \\
  8 &  64.13 & -81.67 & 16.12 $\pm$ 0.01 & 16.18 $\pm$ 0.01 &  15.52
$\pm$  0.01 &    15.14 $\pm$ 0.01 &    14.82 $\pm$   0.01 & \\
\end{tabular}
\end{minipage}
\caption[]{The magnitudes and offset from the OT in arc seconds of the
eight comparison stars used. 
The error listed is the quadratic average
of the measurement error (Poisson noise) and the error originating
from measuring the OT with respect to several reference stars. Not
included is the calibration error, which we estimate to be 0.10. The
V, R and I band images were calibrated using JKT images of the Landolt
PG1528+062 and SA98 fields$^{47}$.  The Landolt star SA104-335, imaged
on Feb. 3.21 UT with the JKT, was used for the U band calibration. We
also calibrated the GRB field in B, V, R and I with images taken with
the KPNO 0.9m, around Feb. 29.07 UT. The V, R and I calibration agrees
within 4 percent for the brighter reference stars with the JKT
values. We adopt the V, R and I calibration of the JKT and the B of
the KPNO 0.9m. For the H band the standard star FS21 (Ref. 48) was
observed during the two nights the OT was detected.}
\end{sidetable}

\end{document}